\documentclass{article}
\usepackage{spconf,amsmath,graphicx}

\usepackage{enumitem}
\setlist{nosep, leftmargin=14pt}

\usepackage{mwe} 

\usepackage{wrapfig}

\title{A Novel Deep Learning Architecture by Integrating Visual Simultaneous Localization and Mapping (vSLAM) into CNN for Real-time Surgical Video Analysis}
\name{Ella Lan}
\address{The Harker School\\
	500 Saratoga Avenue, San Jose, CA 95129}
\begin{document}
%
\maketitle
\begin{abstract}
Seven million people suffer surgical complications each year, but with sufficient surgical training and review, 50\% of these complications could be prevented. To improve surgical performance, existing research uses various deep learning (DL) technologies including convolutional neural networks (CNN) and recurrent neural networks (RNN) to automate surgical tool and workflow detection. However, there is room to improve accuracy; real-time analysis is also minimal due to the complexity of CNN. In this research, a novel DL architecture is proposed to integrate visual simultaneous localization and mapping (vSLAM) into Mask R-CNN. This architecture, vSLAM-CNN (vCNN), for the first time, integrates the best of both worlds, inclusive of (1) vSLAM for object detection, by focusing on geometric information for region proposals, and (2) CNN for object recognition, by focusing on semantic information for image classification, combining them into one joint end-to-end training process. This method, using spatio-temporal information in addition to visual features, is evaluated on M2CAI 2016 challenge datasets, achieving the state-of-the-art results with 96.8 mAP for tool detection and 97.5 mean Jaccard score for workflow detection, surpassing all previous works, and reaching a 50 FPS performance, 10x faster than the region-based CNN. A region proposal module (RPM) replaces the region proposal network (RPN) in Mask R-CNN, accurately placing bounding boxes and lessening the annotation requirement. Furthermore, a Microsoft HoloLens 2 application is developed to provide an augmented reality (AR)-based solution for surgical training and assistance.
\end{abstract}
\begin{keywords}
VSLAM, CNN, Deep Learning, Augmented Reality (AR), Computer Vision
\end{keywords}
\section{Introduction}
\label{sec:intro}

\begin{figure}[htb]
\begin{minipage}[b]{1.0\linewidth}
  \centering
  \centerline{\includegraphics[width=0.8\linewidth]{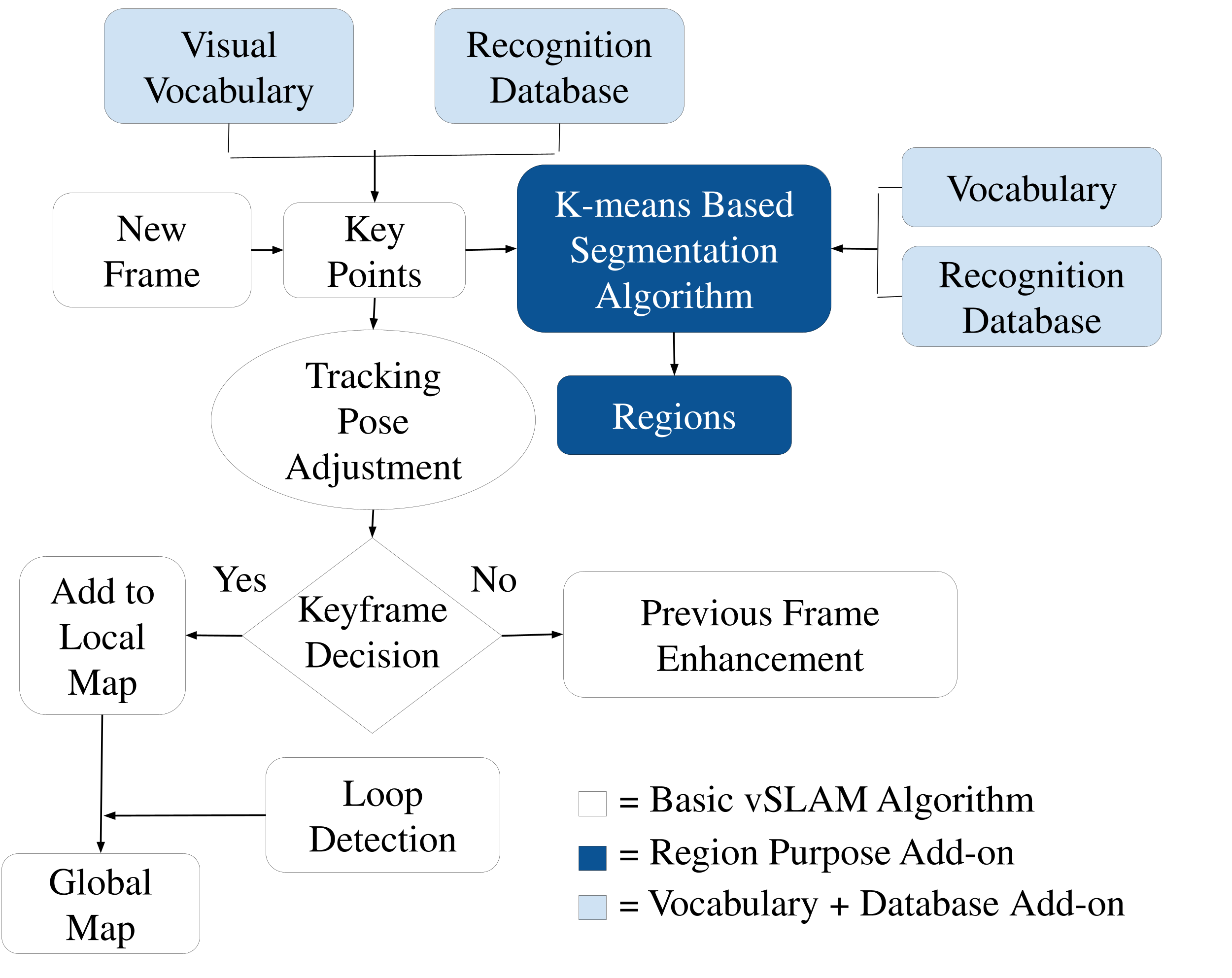}}
\end{minipage}
\caption{The Visual Simultaneous Localization and Mapping (vSLAM) Flow. The main benefits of vSLAM are the key points (the point clouds) that are input into the region proposal module (RPM) for region calculation purpose.}
\label{fig:res}
\end{figure}

Each year, roughly seven million people suffer complications after surgery, but half of those complications could be prevented \cite {alkire2015global}. Effective surgical video review has proved to enhance surgical performance. Recognition of surgical tools and workflows is essential to automate surgical review, monitor surgical processes, and support surgeon decision-making during operations.

Multiple studies have used a variety of neural networks to gather visual features from surgical videos and perform  tool and workflow detection, ranging from challenges such as the M2CAI16 Challenge \cite {twinanda2016endonet} to independent reviews \cite {bouget2017vision} and studies. Twinanda et al. \cite {twinanda2016endonet} extracted visual features via AlexNet convolutional neural networks (CNN); Shvets et al. \cite {shvets2018automatic} performed robotic instrument semantic segmentation via deep learning; Bodenstedt et al. \cite {bodenstedt2018real} calculated bounding boxes via the random forest algorithm. These studies use visual features to detect tool and workflow. However, spatio-temporal information has not been fully used in existing works. Due to the standardization of surgical videos and the U.S. Food and Drug Administration (FDA) protocols regarding the tool usage and order sequence in each workflow \cite {gholinejad2019surgical}, spatio-temporal information can become valuable, potentially improving the model performance.

This paper presents a novel architecture that integrates visual simultaneous localization and mapping (vSLAM) \cite {taketomi2017visual} \cite {mur2015orb} with CNN, named vSLAM-CNN (vCNN). SLAM technology \cite {bailey2006simultaneous} has been applied broadly in the field of robots and self-driving cars but rarely in analysis of surgical videos. The visual SLAM framework, depicted in Figure 1, utilizes computer vision to calculate the position and orientation of a device with respect to its surroundings. And at the same time, vSLAM maps the environment using only visual inputs from a camera for object detection and localization. Under vCNN, a vSLAM-based region proposal module (RPM) is created to replace region proposal network (RPN) \cite {ren2015faster}; it utilizes spatio-temporal information to generate region set proposals from maps created by vSLAM. This architecture improves upon the accuracy in previous studies for both tool and workflow detection by incorporating geometric information and semantic information into the training process. The vCNN also substantially improves the model’s prediction speed through vSLAM’s ability to analyze surgical videos in real-time. Due to the past successes of CNN in object recognition and image classification, the architecture is built atop Mask R-CNN \cite {he2017mask}, utilizing its region of interest (ROI) Align on extracted features and proposed regions.

\section{Related Works}
\label{sec:related}

As can be seen in Figure 2, most of the previous research \cite {sahu2016tool} \cite {jin2018tool} \cite {he2016deep} \cite {jo2019robust} \cite {cadene2016m2cai} \cite {al2018monitoring} \cite {yu2018learning} \cite {nwoye2019weakly} has focused on one genre of information: visual features, spatial information, or temporal information. Studies that use visual features rely on CNN via the fine-tuning approach. For instance, Twinanda et al. \cite {twinanda2016endonet} used fine-tuning CNN to create their own EndoNet architecture, which was the first to use CNN for multiple recognition tasks; in this way, they were able to design a CNN architecture that jointly performs the workflow recognition and tool presence detection tasks using only visual features. Despite their state-of-art results, Twinanda et al. \cite {twinanda2016endonet} were unable to explore localization tracking for tools due to a lack of datasets with spatial bounds. Previous research indicated spatial information has the potential to increase the accuracy of tool detection. Jin et al. \cite {jin2018tool} created a new dataset called m2cai16-tool-locations, which extends the m2cai16-tool dataset (the tool training dataset given by the M2CAI16 Challenge) by adding manual spatial bound annotations around the tools. Jo et al. \cite {jo2019robust} applied YOLO9000 to generate the bounding boxes. Additionally, workflow detection accuracy can benefit from time sequence and temporal information as proved by studies, ranging from hidden Markov model (HMM) \cite {cadene2016m2cai} to recurrent neural networks (RNNs). For instance, Yu et al. \cite {yu2018learning} tracked the time sequence of the surgical video using long short-term memory (LSTM), and Nwoye et al. \cite {nwoye2019weakly} applied convolutional LSTM using temporal information in addition to visual features. These studies suggest that time sequence is a prominent feature in the analysis of phase detection.

\begin{figure}[htb]
\begin{minipage}[b]{0.9\linewidth}
  \centering
  \centerline{\includegraphics[width=0.9\linewidth]{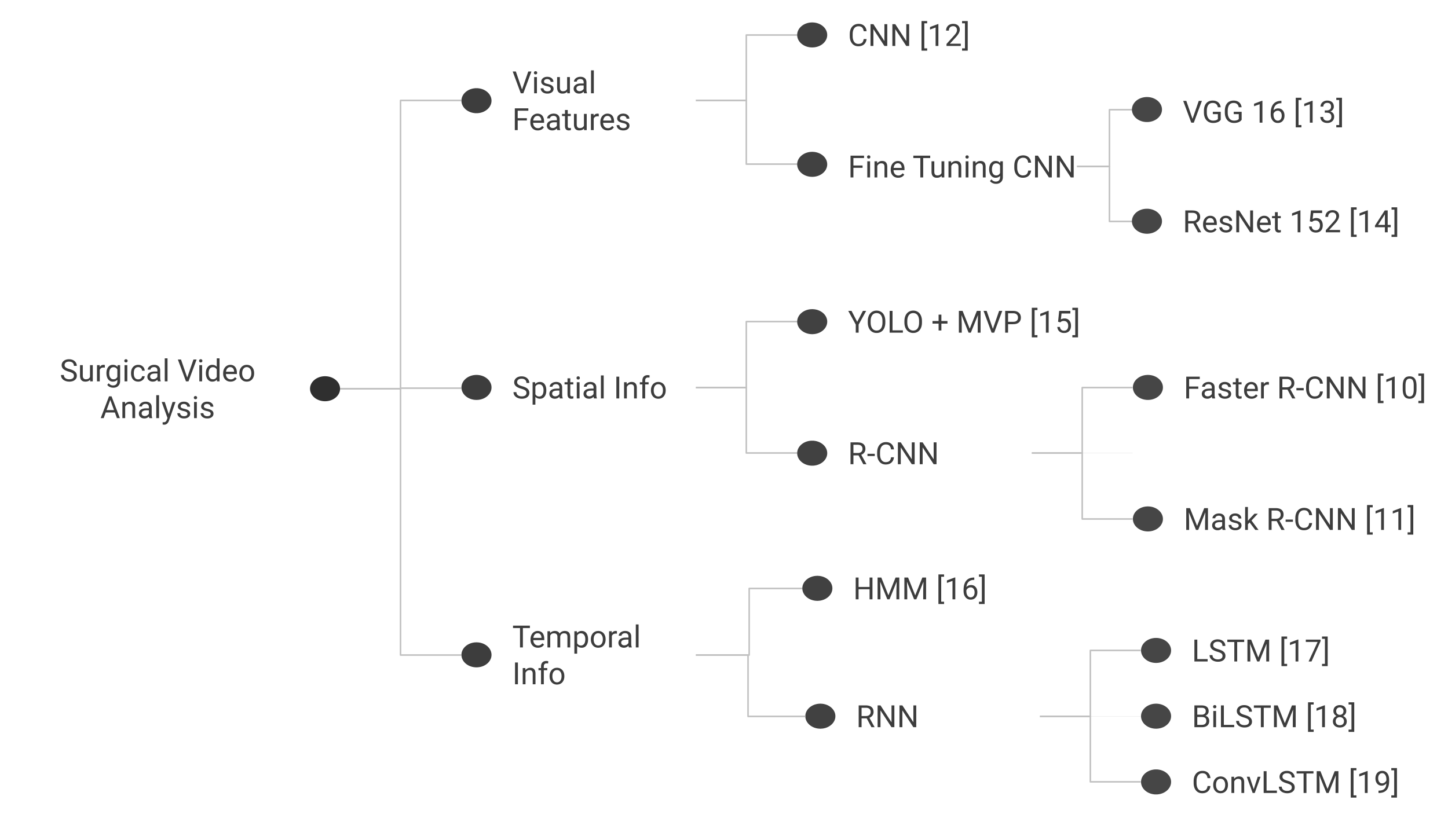}}
\end{minipage}
\caption{The Relationship Between Visual Features, Spatial Information, and Temporal Information When Performing Surgical Video Analysis. This figure summarizes the technologies used in previous research and depicts the framework that was used to develop the concept of creating an architecture that would use visual features and spatio-temporal information by integrating vSLAM into CNN.}
\end{figure}

In summary, previous works have yielded a range of results by using a variety of technologies. However, to maximize the impact of tool and workflow detection on real-world applications, methods yielding accurate prediction and real-time detection speed become critical.

\begin{figure}[htb]
\begin{minipage}[b]{1.0\linewidth}
  \centering
  \centerline{\includegraphics[width=0.8\linewidth]{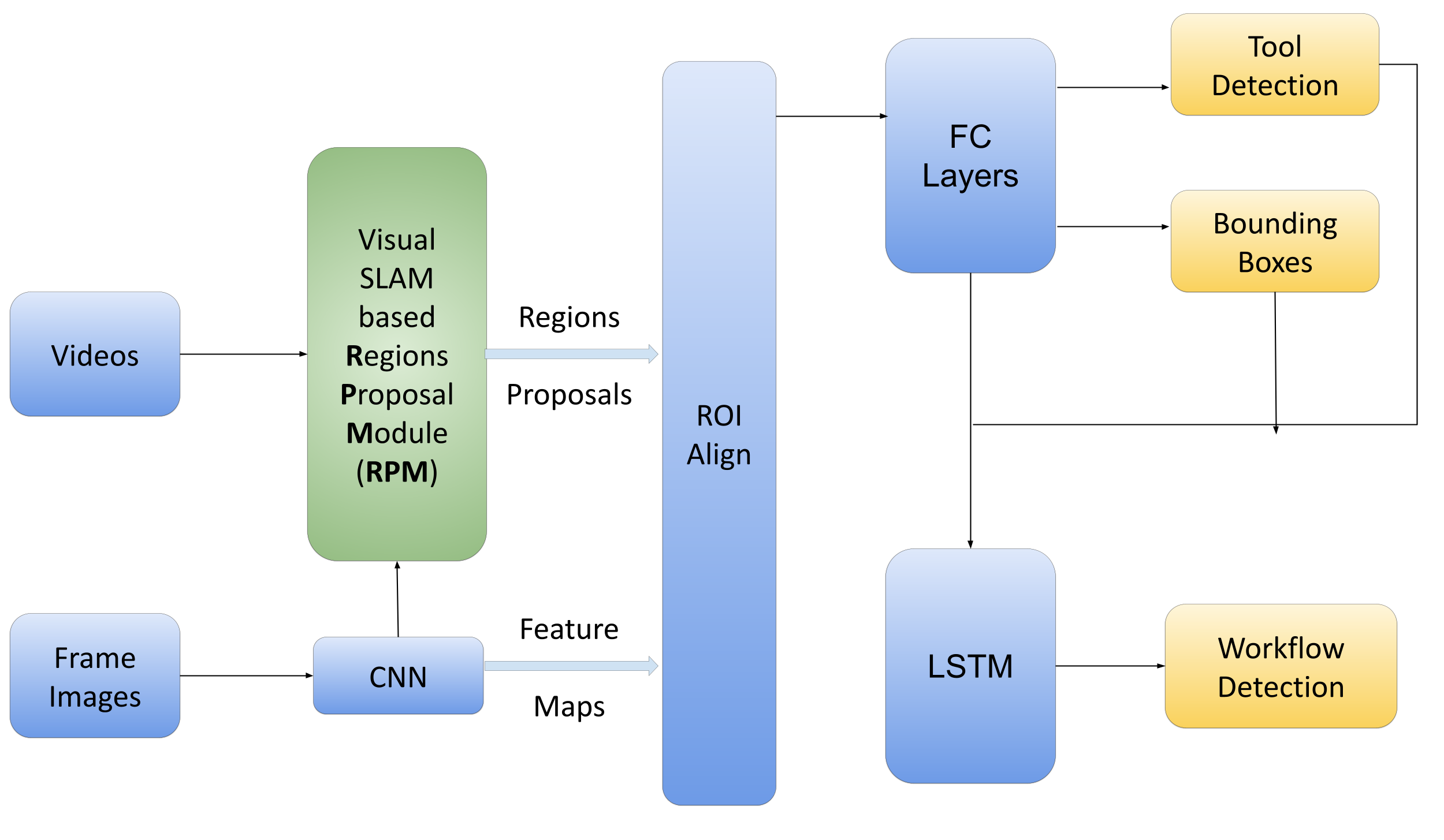}}
\end{minipage}
\caption{The Overall Visual SLAM Convolutional Neural Networks (vCNN) Architecture.}
\end{figure}

Different from other studies, this paper presents a deep learning architecture to perform an end-to-end training for both visual features and spatio-temporal information by integrating vSLAM into Mask R-CNN. This approach would not only improve the model’s performance speed through vSLAM’s instantaneous localization and mapping, but also achieve higher accuracy for both tool and workflow detection, with a lessened requirement of the training data annotations.

\section{Methodology}
\label{sec:method}

The complete pipeline of the proposed approach is depicted in Figure 3. Mask R-CNN is chosen as a backbone due to its success with object classification and ROI Align architecture. The key innovation is to create the visual SLAM-based region proposal module (RPM) to generate region proposals through vSLAM 3D mapping, replacing region proposal network (RPN). From there, region proposals, ROI, are obtained and trained alongside the feature maps through a fine-tuning process. After detecting the tools, the fully connected layers are input into the recurrent neural networks LSTM for workflow detection. The details of this approach are explained as follows.

\subsection{vSLAM-Based Region Proposal Module (RPM)}
\label{ssec:rpm}

RPM is a vSLAM-based module. Through openvslam \cite {sumikura2019openvslam}, an open source framework for visual SLAM (ORB-based \cite {mur2015orb}), 3D maps are generated from the videos captured by monocular cameras, and key points are extracted from the key frames on the maps. Then, region proposals are created through K-means clustering among the key points, as seen in Figure 4. In addition, since these frame images are standardized in size, the typical nine anchors are also created from the centroid of each cluster.

\begin{wrapfigure}{l}{0.25\textwidth}
\includegraphics[width=0.9\linewidth]{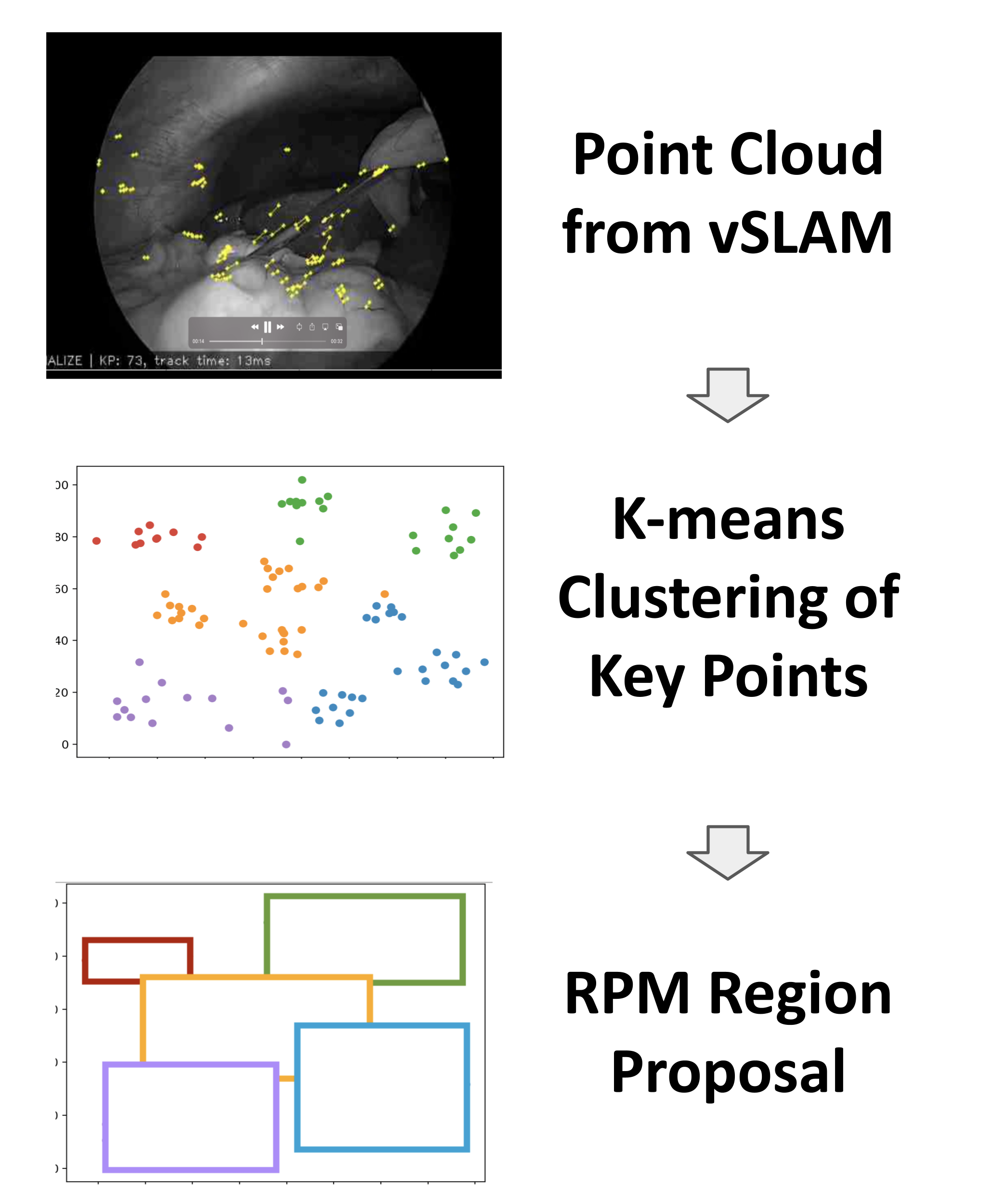}
\caption{An Illustration of Region Generation via K-means.}
\end{wrapfigure}

The surgical videos have unique settings. Their complexity includes dark and blurring surroundings, camera lenses covered by blood, objects overlapping or partially blocked by human organs, small movements in a narrow environment, etc. The RPM architecture is based on the main assumption that through understanding the constant localization of the tools through 3D maps, the proposal regions will accurately capture and present the bounding boxes for the tools. A new set of region generation algorithms has been developed to allow vSLAM to generate sufficient point clouds for surgical videos and to extract the tools location information from 3D maps. 

\begin{itemize}

\item Optimize key frame generation by expanding the time window to utilize more spatio-temporal information to enrich key points’ coverage on tools. Expand the key frame selection from the current frame to include all the key frames within the previous five-second range to generate more regions for the current targeted frame.

\item Adjust and apply the different K values in K-Means, including 2, 3, 4, 5, 6, for each frame, to create broader region proposals.

\item Experiment with additional nine anchors (standard boxes in Faster-RCNN \cite {ren2015faster}) on each centroids from K-means clustering with additional regions.

\end{itemize}

In addition, the different configurations in openvslam, including maximum key points, scale factor, the number of levels, et cetera, are explored to allow vSLAM to generate point clouds on cholecystectomy videos.

\subsection{Mask R-CNN}
\label{ssec:cnn}

\begin{figure}[htb]
\begin{minipage}[b]{1.0\linewidth}
  \centering
  \centerline{\includegraphics[width=0.9\linewidth]{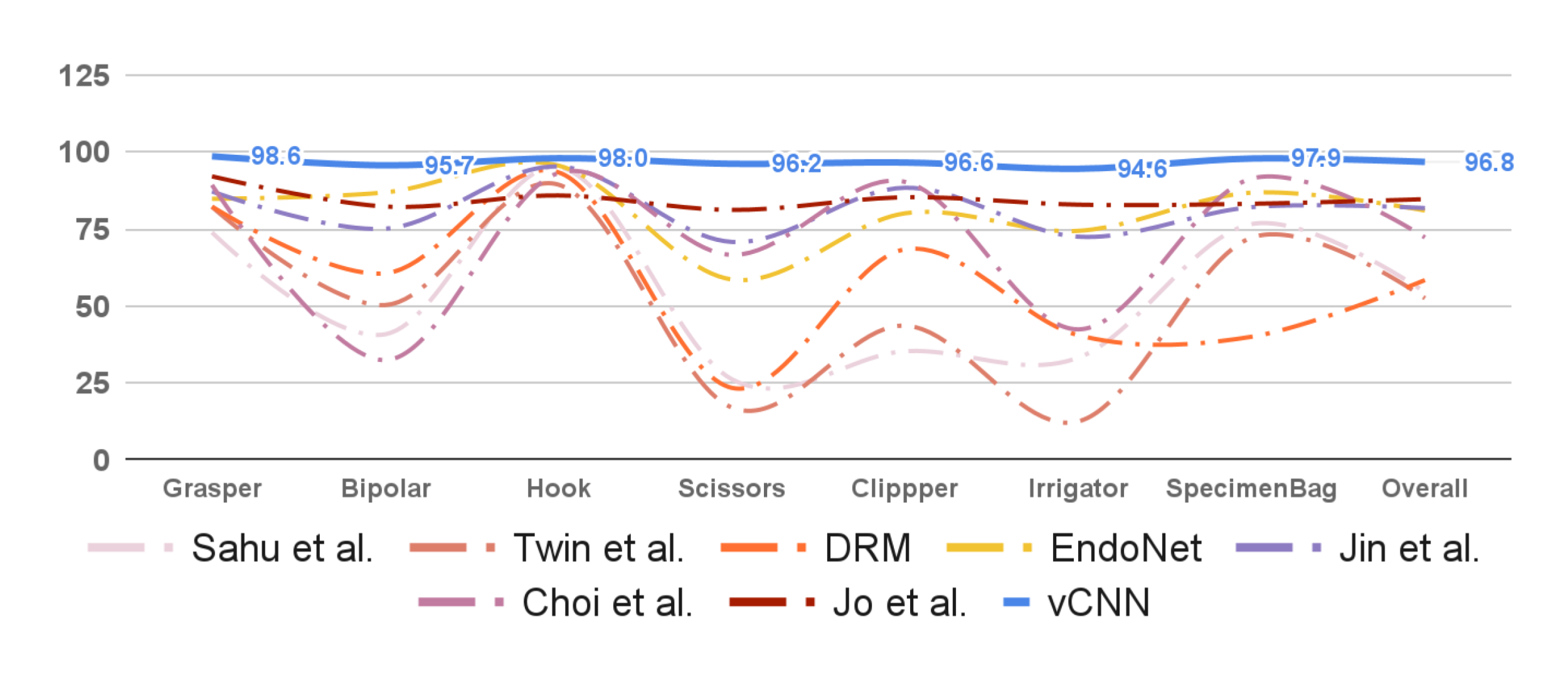}}
\end{minipage}
\caption{The Mean Average Precision (mAP) Comparison of vSLAM-CNN (vCNN) to Previous Studies for Surgical Tool Detection. The vCNN performance achieves the state-of-the-art results.}
\end{figure}

To allow RPM to replace RPN, a technology deeply ingrained inside Faster R-CNN and carried along into the architecture of Mask R-CNN, the first step was to code directly in the open source deep learning learning (DL) framework detectron2 \cite {wu2019detectron2}, and to override its layers forwarding logic to rebuild data flow through the entire pipeline. Although the mask mechanism of Mask R-CNN is not used in this research, the reason for choosing Mask R-CNN is to use its ROI Align layer to retrieve regional proposals from RPM to align with the feature maps from CNN.

\subsection{Fine-Tuning}
\label{ssec:ft}

By leveraging transfer-learning, the fine-tuning approach is used with Mask R-CNN. After loading in the pretrained architecture and weights of the ResNet101 and feature pyramid networks (FPN) \cite {wu2019detectron2}, the model was trained by using custom tools/workflow COCO datasets generated from the original M2CAI16 training data \cite {twinanda2016endonet} \cite {jin2018tool}.

\subsection{Long Short-Term Memory (LSTM) for Workflow}
\label{ssec:lstm}

After the fully connected layers classify the tools, the final tool classification, the created bounding boxes, and the fully connected layer itself, are fed into the LSTM, an extra recurrent neural network (RNN) layer that uses temporal information to predict the workflows.

\section{Experimental Setup and Datasets}
\label{ssec:datasets}

All datasets used in this study consist of cholecystectomy surgical videos collected from the University Hospital of Strasbourg in France. The m2cai16-tool is a dataset that was released for the M2CAI 2016 Tool Presence Detection Challenge \cite {twinanda2016endonet}. The m2cai16-workflow is a dataset that was released for the M2CAI 2016 Workflow Presence Detection Challenge. Another dataset used in the development of the V-CNN architecture is the m2cai16-tool-locations. This is a dataset created by Jin et al. for their study \cite {jin2018tool} and extends upon the original m2cai16-tool by manually adding spatial bounds around the tools. Though it is not essential for spatial bounds of locations to be known because of the replacement of RPN with vSLAM-based RPM, it was utilized to train the vCNN fine-tuning process because of the presumption that locating the tools could potentially increase the accuracy of the overall model. Each surgical video frame is labeled with the certain workflow phase the surgery is in at that time.

\section{Results and Discussions}
\label{sec:results}

\begin{wrapfigure}{l}{0.25\textwidth}
\includegraphics[width=0.9\linewidth]{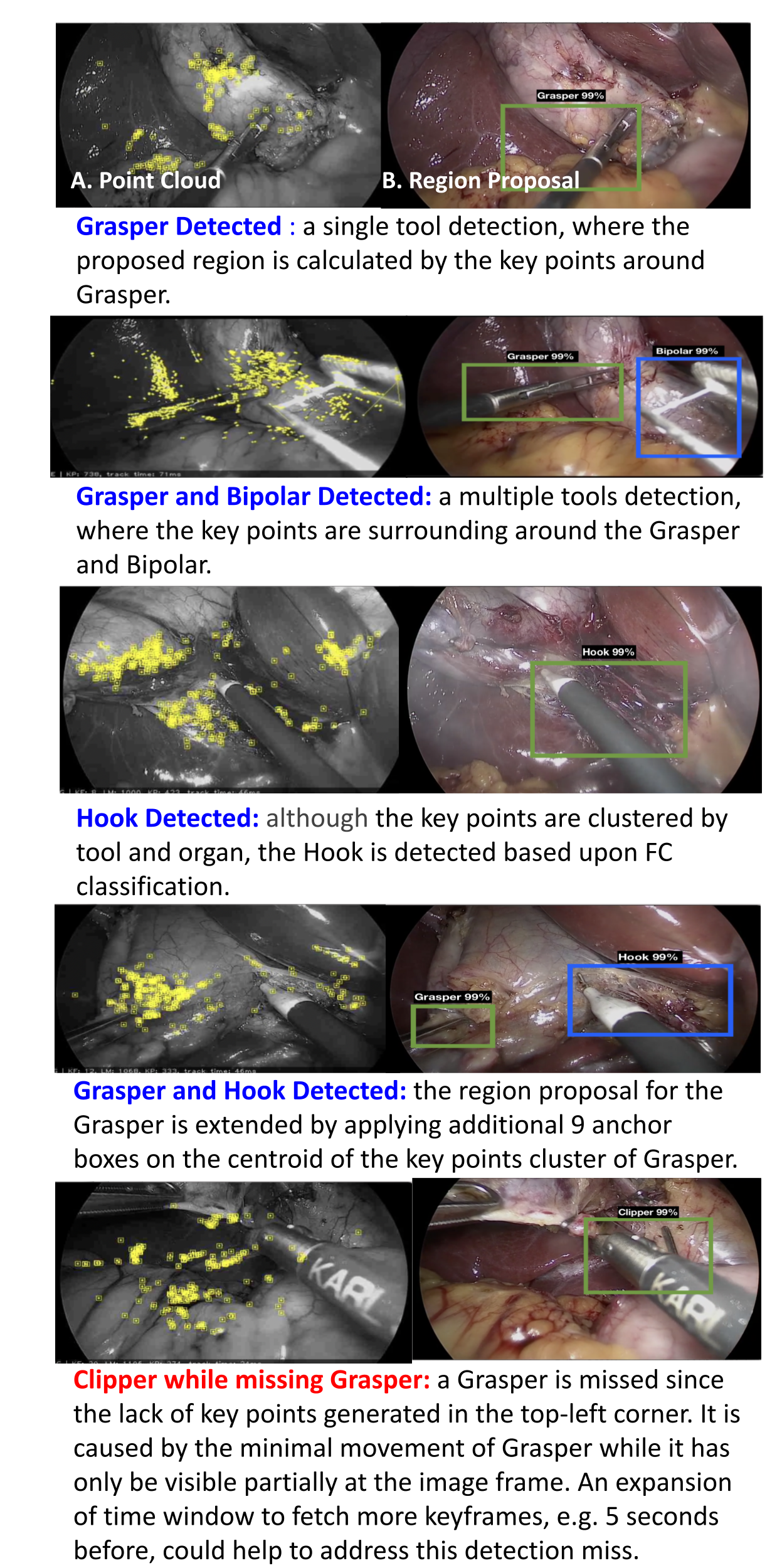}
\caption{Samples of Correct and Incorrect Tool Detection Cases.}
\end{wrapfigure}

The vCNN model generates region proposals by understanding the constant localization from vSLAM 3D maps, and providing DL models with more information to capture and present bounding boxes. By comparing the model’s tool and workflow detection on the blind testing video frames with the ground truth, vCNN's performance is evaluated. The vCNN achieves the state-of-the-art results of 96.8 mAP for tool detection (Figure 5) and 97.5 Jaccard score for workflow detection. It uses feature maps via CNN, regions via RPM, and sequences via LSTM in an end-to-end training process, in which visual features, spatial information, and temporal information are used in a respective manner. 

As indicated in Figure 5, the vCNN generates accurate tool detection, surpassing prior studies by (1) optimizing region generation through the RPM algorithms and (2) fully utilizing spatio-temporal information. And, unlike other studies that have a wide range of detection accuracies for different tools, the mAP of all the tools detection by vCNN is in a more compact range, reflecting the universality of this model and its stable performance under the innovative vCNN architecture. Furthermore, as depicted in Figure 6, an in-depth analysis is performed for both correct and incorrect detection.

Most of R-CNN-based algorithms have a 5 FPS limit due to the RPN complexity; meanwhile, random forests and YOLO directly generates bounding boxes to aim real-time speed (25+ FPS). The vCNN model in run-time takes a total 20 ms, on average, for the blind test prediction, benefited by the instantaneous localization and mapping capability of vSLAM, reaching a real-time speed of 50 FPS.

By understanding real-time surgical videos, the proposed architecture has numerous potential medical applications, such as the automatic indexing of surgical video databases, the monitoring of surgical processes, the alerting of an upcoming complication, the optimization of real-time operating room scheduling, etc. 

The real-time top performance of vCNN, via augmented reality (AR), could be utilized in providing surgical training and even in assisting real-time decisions. To prove the concept, A Microsoft HoloLens 2 application is created to provide real-time surgical (video) analysis via AR as a prototype, presenting a new capability in healthcare applications, including educational training and clinical practices. This application provides a training simulator to interact with surgery procedures in real-time AR, which should help trainees to develop decision-making abilities within weeks rather than months to years from long clinical practice; it is applicable to remote learning and coaching as well.

\section{Conclusions}
\label{sec:contribution}

This paper proposed a deep learning architecture that utilizes visual features and spatio-temporal information gathered from videos. This method is based on a novel architecture called vCNN, which integrates a vSLAM-based RPM with CNN. It demonstrates that connecting visual features, spatial information, and temporal information for the deep learning training can increase the accuracy of tool detection to 96.8 mAP and workflow detection to 97.5 mean Jaccard on the surgical video datasets of M2CAI 2016 challenge. The vSLAM process is capable of generating 3D maps from the surgery videos recorded by a monocular camera. RPM can effectively generate region proposals from vSLAM 3D maps through constant tracking and localization, offering a high performance solution to capture and present bounding boxes in real-time, replacing RPN with comparable high accuracy and removing the ground truth requirements for bounding box annotations in the training datasets. 

Future work includes continuing the expansion of research to higher-quality video recordings via stereo or RGB-D cameras, for better vSLAM/RPM performance, while also exploring general object detection in computer vision for biomedical imaging.

\bibliographystyle{IEEEbib}
\bibliography{strings,refs}

\begin{thebibliography}{10}

\bibitem{alkire2015global}
Blake~C Alkire, Nakul~P Raykar, Mark~G Shrime, Thomas~G Weiser, Stephen~W
  Bickler, John~A Rose, Cameron~T Nutt, Sarah~LM Greenberg, Meera Kotagal,
  Johanna~N Riesel, et~al.,
\newblock ``Global access to surgical care: a modelling study,''
\newblock {\em The Lancet Global Health}, vol. 3, no. 6, pp. e316--e323, 2015.

\bibitem{twinanda2016endonet}
Andru~P Twinanda, Sherif Shehata, Didier Mutter, Jacques Marescaux, Michel
  De~Mathelin, and Nicolas Padoy,
\newblock ``Endonet: a deep architecture for recognition tasks on laparoscopic
  videos,''
\newblock {\em IEEE transactions on medical imaging}, vol. 36, no. 1, pp.
  86--97, 2016.

\bibitem{bouget2017vision}
David Bouget, Max Allan, Danail Stoyanov, and Pierre Jannin,
\newblock ``Vision-based and marker-less surgical tool detection and tracking:
  a review of the literature,''
\newblock {\em Medical image analysis}, vol. 35, pp. 633--654, 2017.

\bibitem{shvets2018automatic}
Alexey~A Shvets, Alexander Rakhlin, Alexandr~A Kalinin, and Vladimir~I
  Iglovikov,
\newblock ``Automatic instrument segmentation in robot-assisted surgery using
  deep learning,''
\newblock in {\em 2018 17th IEEE International Conference on Machine Learning
  and Applications (ICMLA)}. IEEE, 2018, pp. 624--628.

\bibitem{bodenstedt2018real}
Sebastian Bodenstedt, Antonia Ohnemus, Darko Katic, Anna-Laura Wekerle, Martin
  Wagner, Hannes Kenngott, Beat M{\"u}ller-Stich, R{\"u}diger Dillmann, and
  Stefanie Speidel,
\newblock ``Real-time image-based instrument classification for laparoscopic
  surgery,''
\newblock {\em arXiv preprint arXiv:1808.00178}, 2018.

\bibitem{gholinejad2019surgical}
Maryam Gholinejad, Arjo J.~Loeve, and Jenny Dankelman,
\newblock ``Surgical process modelling strategies: which method to choose for
  determining workflow?,''
\newblock {\em Minimally Invasive Therapy \& Allied Technologies}, vol. 28, no.
  2, pp. 91--104, 2019.

\bibitem{taketomi2017visual}
Takafumi Taketomi, Hideaki Uchiyama, and Sei Ikeda,
\newblock ``Visual slam algorithms: a survey from 2010 to 2016,''
\newblock {\em IPSJ Transactions on Computer Vision and Applications}, vol. 9,
  no. 1, pp. 1--11, 2017.

\bibitem{mur2015orb}
Raul Mur-Artal, Jose Maria~Martinez Montiel, and Juan~D Tardos,
\newblock ``Orb-slam: a versatile and accurate monocular slam system,''
\newblock {\em IEEE transactions on robotics}, vol. 31, no. 5, pp. 1147--1163,
  2015.

\bibitem{bailey2006simultaneous}
Tim Bailey and Hugh Durrant-Whyte,
\newblock ``Simultaneous localization and mapping (slam): Part ii,''
\newblock {\em IEEE robotics \& automation magazine}, vol. 13, no. 3, pp.
  108--117, 2006.

\bibitem{ren2015faster}
Shaoqing Ren, Kaiming He, Ross Girshick, and Jian Sun,
\newblock ``Faster r-cnn: Towards real-time object detection with region
  proposal networks,''
\newblock {\em Advances in neural information processing systems}, vol. 28, pp.
  91--99, 2015.

\bibitem{he2017mask}
Kaiming He, Georgia Gkioxari, Piotr Doll{\'a}r, and Ross Girshick,
\newblock ``Mask r-cnn,''
\newblock in {\em Proceedings of the IEEE international conference on computer
  vision}, 2017, pp. 2961--2969.

\bibitem{sahu2016tool}
Manish Sahu, Anirban Mukhopadhyay, Angelika Szengel, and Stefan Zachow,
\newblock ``Tool and phase recognition using contextual cnn features,''
\newblock {\em arXiv preprint arXiv:1610.08854}, 2016.

\bibitem{jin2018tool}
Amy Jin, Serena Yeung, Jeffrey Jopling, Jonathan Krause, Dan Azagury, Arnold
  Milstein, and Li~Fei-Fei,
\newblock ``Tool detection and operative skill assessment in surgical videos
  using region-based convolutional neural networks,''
\newblock in {\em 2018 IEEE Winter Conference on Applications of Computer
  Vision (WACV)}. IEEE, 2018, pp. 691--699.

\bibitem{he2016deep}
Kaiming He, Xiangyu Zhang, Shaoqing Ren, and Jian Sun,
\newblock ``Deep residual learning for image recognition,''
\newblock in {\em Proceedings of the IEEE conference on computer vision and
  pattern recognition}, 2016, pp. 770--778.

\bibitem{jo2019robust}
Kyungmin Jo, Yuna Choi, Jaesoon Choi, and Jong~Woo Chung,
\newblock ``Robust real-time detection of laparoscopic instruments in robot
  surgery using convolutional neural networks with motion vector prediction,''
\newblock {\em Applied Sciences}, vol. 9, no. 14, pp. 2865, 2019.

\bibitem{cadene2016m2cai}
Remi Cadene, Thomas Robert, Nicolas Thome, and Matthieu Cord,
\newblock ``M2cai workflow challenge: convolutional neural networks with time
  smoothing and hidden markov model for video frames classification,''
\newblock {\em arXiv preprint arXiv:1610.05541}, 2016.

\bibitem{al2018monitoring}
Hassan Al~Hajj, Mathieu Lamard, Pierre-Henri Conze, B{\'e}atrice Cochener, and
  Gwenol{\'e} Quellec,
\newblock ``Monitoring tool usage in surgery videos using boosted convolutional
  and recurrent neural networks,''
\newblock {\em Medical image analysis}, vol. 47, pp. 203--218, 2018.

\bibitem{yu2018learning}
Tong Yu, Didier Mutter, Jacques Marescaux, and Nicolas Padoy,
\newblock ``Learning from a tiny dataset of manual annotations: a
  teacher/student approach for surgical phase recognition,''
\newblock {\em arXiv preprint arXiv:1812.00033}, 2018.

\bibitem{nwoye2019weakly}
Chinedu~Innocent Nwoye, Didier Mutter, Jacques Marescaux, and Nicolas Padoy,
\newblock ``Weakly supervised convolutional lstm approach for tool tracking in
  laparoscopic videos,''
\newblock {\em International journal of computer assisted radiology and
  surgery}, vol. 14, no. 6, pp. 1059--1067, 2019.

\bibitem{sumikura2019openvslam}
Shinya Sumikura, Mikiya Shibuya, and Ken Sakurada,
\newblock ``Openvslam: a versatile visual slam framework,''
\newblock in {\em Proceedings of the 27th ACM International Conference on
  Multimedia}, 2019, pp. 2292--2295.

\bibitem{wu2019detectron2}
Yuxin Wu, Alexander Kirillov, Francisco Massa, Wan-Yen Lo, and Ross Girshick,
\newblock ``Detectron2. 2019,''
\newblock {\em URL https://github. com/facebookresearch/detectron2}, vol. 2,
  no. 3, 2019.

\end{thebibliography}

\end{document}